\documentclass[dvips]{article}

\usepackage{icrc2011}

\title{Performance of the MAGIC Stereo System}

\newcommand{\etal}{\MakeLowercase{\textit{et al. }}} 
\shorttitle{E. Carmona \etal for the MAGIC Collaboration}

\authors{E. Carmona$^{1}$, J. Sitarek$^{2}$, P. Colin$^{3}$, M. Doert$^{4}$,
  S. Klepser$^{5}$, S. Lombardi$^{6}$, M. L\'opez$^{7}$, A. Moralejo$^{5}$,
  S. Pardo$^{7}$, V. Scalzotto$^{6}$, R. Zanin$^{5}$ 
  for the MAGIC Collaboration}
\afiliations{$^1$Centro de Investigaciones Energ\'eticas,
  Medioambientales y Tecnol\'ogicas, Madrid, Spain\\
 $^2$Division of Astrophysics, University of Lodz, Lodz, Poland \\
 $^3$Max-Planck-Institut f\"ur Physik, Munich, Germany\\
 $^4$Universit\"at Dortmund, Dortmund, Germany  \\
 $^5$Institut de F\'isica d'Altes Energies, Barcelona, Spain \\
 $^6$Dipartimento di Fisica, Universit\`a di Padova and INFN sez. di  Padova, Padova, Italy  \\
 $^7$Universidad Complutense, Madrid, Spain 
 }
\email{emiliano.carmona@ciemat.es}

\abstract{ MAGIC is a system of two Imaging Atmospheric Cherenkov
  Telescopes sensitive above $\sim$60~GeV, and located on the Canary
  Island of La Palma at the height of 2200 m.a.s.l.  Since Autumn 2009
  both telescopes are working together in stereoscopic mode.  We use
  both Crab Nebula observations and Monte Carlo simulations to
  evaluate the performance of the system.  Advanced stereo analysis
  allows MAGIC to achieve a sensitivity better than 0.8\% of the Crab
  Nebula flux in 50~h of observations in the medium energy range
  (around a few hundred GeV).  At those energies the angular
  resolution is better than 0.07$^\circ$, and the energy resolution is
  as good as 16\%.  We perform also a detailed study of possible
  systematics effects for the MAGIC telescopes.  }


\begin{document}
\maketitle


\section{The MAGIC Telescopes}

The MAGIC experiment is located at a height of 2200 m.a.s.l. on the
Canary island of La Palma. The system is composed of two 17~m Imaging
Atmospheric Cherenkov Telescopes (IACTs)  devoted to the
observation of very high energy (VHE, $>$30~GeV) gamma rays. The
first of the MAGIC telescopes (MAGIC~I) started operations in
2004. MAGIC~II was built some years later allowing stereo observations 
since the autumn of 2009. 

MAGIC~II was constructed like a copy of MAGIC~I with a few
improvements. Both telescopes are built using the same light-weight
carbon-fibre structure. The size of the mirror dish (17~m diameter) and
the camera field of view (3.5$^\circ$) are the same in both
telescopes. The main difference between the telescopes are
their cameras. MAGIC~I camera is composed of 577 hexagonal pixels with
an angular size of $0.1^\circ$ in the inner part of the camera and
$0.2^\circ$ in the outer part. On the contrary, the MAGIC~II camera is composed
of 1039 $0.1^\circ$ hexagonal pixels.

In both telescopes the signals from the PMTs in each pixel are
optically transmitted to the counting house where trigger and
digitisation of the signals take place. The trigger area of the
MAGIC~II camera has a radius of $1.25^\circ$. This is an improvement
with respect to MAGIC~I that has a radius of only
$0.95^\circ$. The signals of both telescopes are digitised using a
frequency of 2 GSamples/s. In the case of MAGIC~I a system based on
multiplexed FADCs is used~\cite{Goebel2007}. For MAGIC~II the readout
system is based on the Domino Ring Sampler 2 (DRS2) chip~\cite{Tescaro}.

Regular observations are performed in stereoscopic mode. Only events
that trigger both telescopes are recorded. The trigger condition
for the individual telescope is the so-called 3NN trigger. Each
telescope must have 3 pixels above their pixel threshold ({\em
  level-0} trigger) in a next-neighbour (NN) topology in order to be a
telescope trigger ({\em level-1} trigger). The stereo trigger makes a
tight time coincidence between both telescopes taking into account the
delay due to the relative position of the telescopes and their
pointing direction. Although the individual telescope trigger rates
are high (several kHz), the stereo trigger rate is in the range of
150-200~Hz with just a few Hz being accidental triggers.

\section{Data Sample}

In order to study the performance of the MAGIC telescopes, we have
used a sample of Crab Nebula data. The Crab Nebula is considered the
standard candle for VHE gamma-ray astronomy since it is the brightest
steady, point-like VHE source in the sky. Although the source seems to
be variable in the GeV energy range (ATel \#2855, \cite{Tavani2011},
\cite{Abdo2011}) neither MAGIC nor VERITAS (ATel \#2967, ATel \#2968)
have observed variability in the VHE flux~\cite{Zanin}.

The data used in  this study was taken between November
2009 and January 2010. After data quality selection, 9~hours of good quality
data were used. These data were taken at low zenith angle
($<30^\circ$) in wobble mode~\cite{Fomin1994} with the source located
at a distance of $0.4^\circ$ from the camera centre. The position of
the source in the camera is rotated $180^\circ$ around the camera
centre every 20 minutes to decrease the systematic errors that could
come from camera inhomogeneities. 

The performance of the MAGIC system have also been studied using
Monte Carlo (MC) data. MC gamma rays with an energy range from 30~GeV
to 30~TeV have been used to study the response of MAGIC to signal
events. In addition, MC protons (30~GeV to 30~TeV), helium (70~GeV to
20~TeV), and electrons (70~GeV to 7~TeV) where used to compare the
cosmic background rates in data with MC predictions.

\section{Stereo Analysis}

The data analysis is done using the MAGIC standard analysis framework
(MARS)~\cite{MoralejoLodz}. The first steps in the stereo data
analysis are done independently for each telescope (calibration, image
cleaning, image parametrisation), as described
in~\cite{Albert2008}\cite{Aliu2009}. After the relevant parameters of
each image are obtained, the stereo analysis is applied, combining the
information obtained from both telescopes. A summary of the recent
advancements introduced in the stereoscopic analysis can be found
in~\cite{icrc_advanced_analysis}. The new stereo parameters {\em Impact}
(reconstructed distance from each telescope to the shower axis in the
plane perpendicular to the shower axis) and {\em MaxHeight}
(reconstructed height of the shower maximum above the telescopes) are
introduced in the data analysis. The
stereoscopic observations allow us to easily obtain these geometrical
parameters that improve the energy reconstruction (mainly due to the
{\em Impact} parameter) and increase the background rejection (mainly
due to the {\em MaxHeight} parameter). At low energies, background
rejection is significantly enhanced by the {\em MaxHeight} because a
large fraction of background events are reconstructed with {\em
  MaxHeight} below 4~km above the telescopes. These are muon events
that now can be efficiently rejected with the stereo system.

The reconstructed direction of the primary particle is another
parameter that can be obtained geometrically in stereo events.  The
crossing point of the main axes of the ellipses that the event forms
in each camera gives the direction of the incoming primary
particle~\cite{Aharonian97}. In MAGIC, however, we combine this
information with the timing information to improve the angular
resolution. For each individual image, the DISP parameter (distance
from the shower image centre to the estimated source position) is
estimated using the multidimensional classification method Random
Forest~\cite{Albert2008c} (RF). We call this method DISP-RF and it is
applied to the images of each telescope. Later, the geometrical
information from the axes crossing point plus the two estimations from
the DISP-RF method are combined to provide a unique source position
estimation or to reject the event if the individual positions do not
agree, which provides additional background rejection (more details
are given in~\cite{icrc_advanced_analysis}~\cite{JSitarek2011}).

The RF method is also used for gamma-hadron separation in
the stereo data. In this case, the information from some selected
individual telescope parameters like {\em Size} (sum of charge in all
pixels surviving the image cleaning), {\em Width} (angular distance
along the minor axis of the image ellipse in the camera) or {\em
  Length} (angular distance along the major axis of the image ellipse)
are combined with the stereo parameters and timing information. The
result of this process is a single parameter called {\em Hadronness}
which is distributed between 0 (gamma-like event) and 1 (hadron-like
event) and it is used in the analysis to select the gamma-like events.
The parameters used to build {\em Hadronness} are source-independent,
so {\em Hadronness} is not biased by the position of the source in the
camera.

Finally the energy of the events is reconstructed using lookup tables
where the size of the image, {\em Impact} and {\em MaxHeight} are used
to obtain an energy estimation of the event with the parameters from
each telescope (more details are given in~\cite{JSitarek2011}). A weighted
mean is calculated with the energy estimation from each telescope to
produce the estimated energy of the event. The tables are filled with
the information of MC gamma rays with the same zenith angle as the
data being analysed.

\section{Performance of the MAGIC Stereo System}

The energy threshold of the MAGIC stereo system (defined as the peak
of the distribution of gamma-ray events after trigger) has been
estimated as 50~GeV for a differential spectral index of -2.6. For the
measured Crab Nebula spectrum the threshold is 60~GeV
(see~\cite{JSitarek2011} for more details). After image cleaning, 
the rate of cosmic ray events obtained from MC simulations at low sizes
agrees with data within 20\%. It shows a larger discrepancy at higher
energies that could come from systematic errors in both the
measurement by BESS~\cite{Yamamoto2007}, which the MC is normalised
with, and/or MAGIC as well as the limited acceptance of the MC
simulations.

\begin{figure}[!t]
  \vspace{5mm} \centering
  \includegraphics[width=0.95\columnwidth]{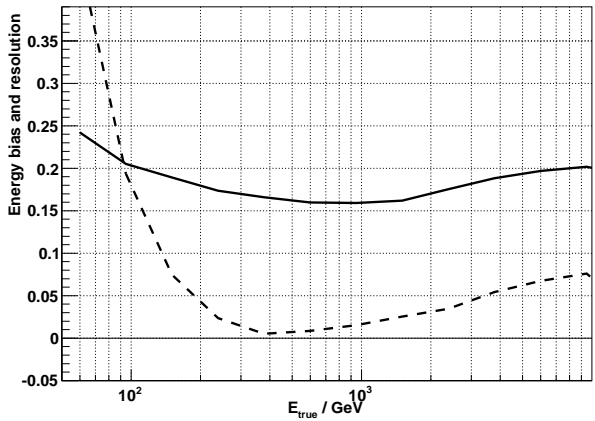}
  \caption{Energy resolution (solid line) and bias (dashed line)
    obtained from MC gamma-ray simulations.}
  \label{Energy}
 \end{figure}

The energy resolution and energy bias obtained with the MAGIC stereo
system are shown in Figure~\ref{Energy}. 
The best value reached is 16\% at a few hundred GeV. At higher
energies it is slightly worse due to limited statistics and events
being truncated at the edge of the camera.

\begin{figure}[!t]
  \vspace{5mm} \centering
  \includegraphics[width=0.95\columnwidth]{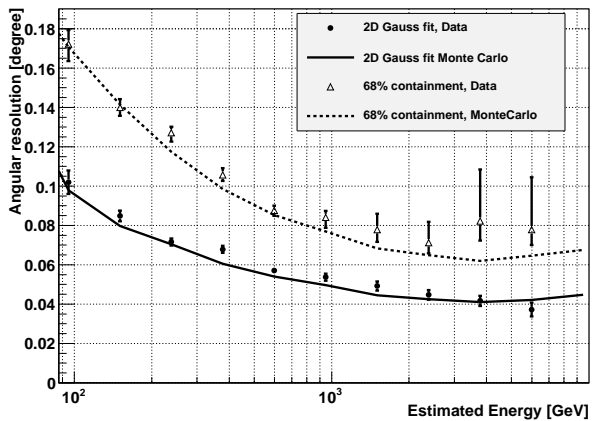}
  \caption{Angular resolution obtained from the Crab data sample
    (points) and from MC simulations (lines) as a function of the
    estimated energy. The angular resolution is shown as the sigma of a 2D
    gaussian fit (solid circles, solid line) or as the 68\%
    containment radius (empty triangles, dashed line).}
  \label{Theta2}
 \end{figure}

The angular resolution obtained from data events and MC simulations is
shown in Figure~\ref{Theta2}. With the use of DISP-RF, an angular
resolution of 0.07$^\circ$ is reached at 300~GeV. MC simulations agree
well with the data and only a small discrepancy is seen at high
energies coming from small imperfections not included in MC simulations that
is visible when the angular resolution is very small. 

\begin{figure}[!t]
  \vspace{5mm} \centering
  \includegraphics[width=0.95\columnwidth]{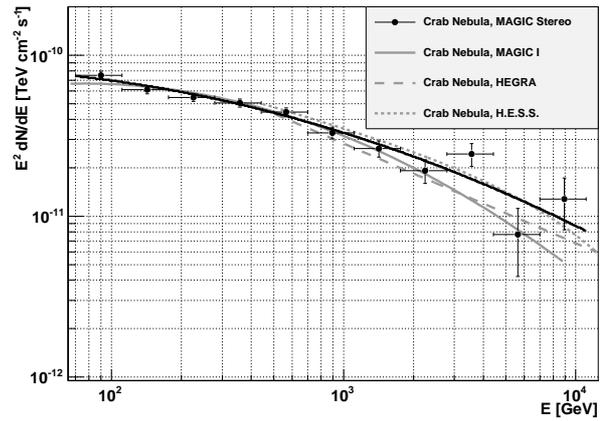}
  \caption{SED of the Crab Nebula obtained by MAGIC stereo (black
    points and line) compared to other experiments~\cite{Albert2008}\cite{Aharonian2004}\cite{Aharonian2006} (grey).}
  \label{FluxSED}
 \end{figure}

The SED obtained from the Crab Nebula data sample is shown in
Figure~\ref{FluxSED}. In the energy range 70~GeV-11~TeV the Crab
Nebula spectrum can be fitted by:

\begin{eqnarray*}
\frac{dN}{dE}=f_0( E/ 300\ \mathrm{GeV}\,) ^{a+b\mathrm{\log}_{10}(\frac{E}{300}\,\mathrm{GeV}\,)}\ 
\mathrm{cm^{-2}s^{-1}TeV^{-1}}
\label{eq_spectrum}
\end{eqnarray*} 
where
$f_0 =( 5.8\pm 0.1_{\mathrm{stat}}) \times 10^{-10}$, 
$a=-2.32\pm 0.02_{\mathrm{stat}}$, 
and $b=-0.13\pm 0.04_{\mathrm{stat}}$. A detailed study of the Crab
Nebula is presented in~\cite{Zanin2011}.

\begin{figure}[!t]
  \vspace{5mm} \centering
  \includegraphics[width=0.95\columnwidth]{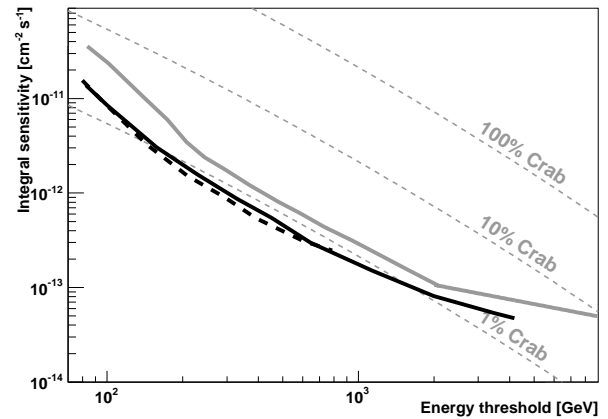}
  \caption{Integral sensitivity of the MAGIC Stereo system, defined as
    the flux above a given energy for which $N_{\rm
      excess}/\sqrt{N_{\rm bgd}}=5$ after 50~h. The solid
    black line shows the values obtained from the Crab data. The
    dashed black line shows the values obtained from MC estimations (MC
    gamma rays and MC backgrounds including protons, helium and
    electrons). In the same figure the MAGIC~I sensitivity is shown as
    a grey solid line. Different fractions of the Crab
    Nebula flux (dashed grey) are also shown.} 
  \label{IntSens}
 \end{figure}

\begin{figure}[!t]
  \vspace{5mm} \centering
  \includegraphics[width=0.95\columnwidth]{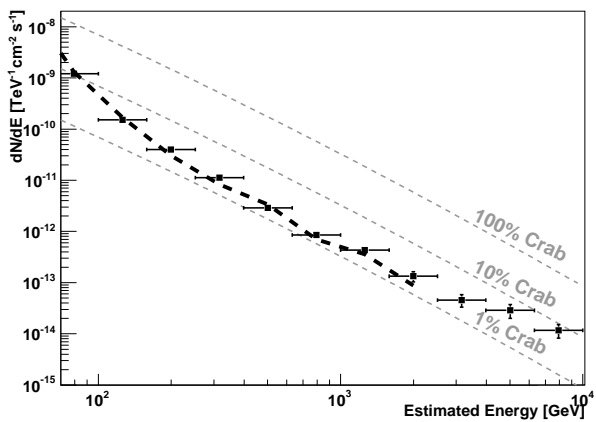}
  \caption{Differential sensitivity of the MAGIC Stereo
    system given by the flux that provides $N_{\rm
      excess}/\sqrt{N_{\rm bgd}}=5$, $N_{\rm excess}>10$ and $N_{\rm
      excess}> 0.05 N_{\rm bgd}$ after 50~h of effective time (black
    points). The result obtained from MC simulations is
   also shown as a dashed black line. }
  \label{DiffSens}
 \end{figure}

The integral sensitivity of the instrument as a function of the energy
is shown in Figure~\ref{IntSens}, while the differential sensitivity
is shown in Figure~\ref{DiffSens}. In both cases there is a good
agreement between the sensitivity estimations from MC and the actual
sensitivity reached with the Crab data sample. The second telescope
and the use of stereo data analysis provide a significant sensitivity
improvement in comparison to MAGIC~I alone. A factor $\sim$2
improvement in significance is achieved at a few hundred GeV and up to
a factor $\sim$3 at lower energies. Values for the integral
sensitivity reached by MAGIC for a Crab Nebula spectrum as a function
of the threshold energy can be found in table~\ref{tab_sens}.

\begin{table*}[t]
\centering
\small
\begin{tabular*}{0.7\textwidth}{@{\extracolsep{\fill}} c|c|c|c|c|c}
$E_{\rm thresh.}$ & $S_{5\sigma, 50\mathrm{h}}$ & $S_{\rm syst}$ & $S_{\rm Li\&Ma, 1Off}$ & $S_{\rm Li\&Ma, 3Off}$ & $S_{5\sigma, 50\mathrm{h}}$  \\
~[GeV] & [\%C.U.] & [\%C.U.] & [\%C.U.] & [\%C.U.]&[$10^{-13}$ $\rm cm^{-2} s^{-1}$] \\\hline\hline
82.5 & $1.99 \pm 0.03$ & $4.7 \pm 0.04$ & $2.84 \pm 0.019$ & $2.314 \pm 0.016$ & $138 \pm 2$ \\ \hline
100 & $1.56 \pm 0.02$ & $2.41 \pm 0.03$ & $2.24 \pm 0.02$ & $1.818 \pm 0.018$ & $84.1 \pm 1.3$ \\ \hline
158 & $1.02 \pm 0.02$ & $1.02 \pm 0.02$ & $1.48 \pm 0.03$ & $1.2 \pm 0.02$ & $30.4 \pm 0.6$ \\ \hline
229 & $0.87 \pm 0.02$ & $0.87 \pm 0.02$ & $1.29 \pm 0.03$ & $1.03 \pm 0.02$ & $15.7 \pm 0.4$ \\ \hline
328 & $0.79 \pm 0.03$ & $0.79 \pm 0.03$ & $1.23 \pm 0.05$ & $0.97 \pm 0.04$ & $8.7 \pm 0.4$ \\ \hline
452 & $0.78 \pm 0.04$ & $0.78 \pm 0.04$ & $1.25 \pm 0.06$ & $0.98 \pm 0.05$ & $5.4 \pm 0.3$ \\ \hline
646 & $0.72 \pm 0.06$ & $0.72 \pm 0.06$ & $1.23 \pm 0.08$ & $0.95 \pm 0.07$ & $3 \pm 0.3$ \\ \hline
1130 & $0.86 \pm 0.06$ & $0.86 \pm 0.06$ & $1.67 \pm 0.07$ & $1.23 \pm 0.06$ & $1.51 \pm 0.11$ \\ \hline
2000 & $1.12 \pm 0.14$ & $1.12 \pm 0.14$ & $2.67 \pm 0.16$ & $1.85 \pm 0.13$ & $0.81 \pm 0.1$ \\ \hline
2730 & $1.5 \pm 0.3$ & $1.58 \pm 0.15$ & $4.3 \pm 0.3$ & $2.8 \pm 0.3$ & $0.64 \pm 0.12$ \\ \hline
3490 & $1.8 \pm 0.4$ & $2.3 \pm 0.3$ & $5.9 \pm 0.5$ & $3.8 \pm 0.4$ & $0.53 \pm 0.11$ \\ \hline
4180 & $2.3 \pm 0.5$ & $2.9 \pm 0.4$ & $7.5 \pm 0.6$ & $4.8 \pm 0.5$ & $0.47 \pm 0.1$ \\ \hline
\end{tabular*}
\caption{
Integral sensitivity obtained with the Crab Nebula data sample above the energy threshold $E_{\rm thresh.}$.
The sensitivity is calculated as $N_{\rm excess}/\sqrt{N_{\rm bgd}}=5$
after 50~h of the effective time ($S_{5\sigma, 50\mathrm{h}}$), or
with additional conditions $N_{\rm excess}>10$, $N_{\rm excess}> 0.05
N_{\rm bgd}$ ($S_{\rm syst}$). The effect of the number of OFF regions
used to compute the background (1 or 3) is shown in $S_{\rm Li\&Ma, 1Off}$ and
$S_{\rm Li\&Ma, 3Off}$ columns respectively. } 
\label{tab_sens}
\end{table*}
%


Finally, the sensitivity for a source located at different distances
from the camera centre has also been investigated. Dedicated
observations of the Crab Nebula were carried out with the source
located at different offset angles between 0.2$^\circ$ and
1.4$^\circ$. The results show that the integral sensitivity obtained above
290~GeV for Crab is almost constant up to an offset angle of
$\sim 0.5^\circ$. For higher distances from the camera centre, the
sensitivity degrades significantly. For a source located at 1$^\circ$ from the camera
centre, the integral sensitivity above 290~GeV degrades by a factor
$\sim$2.

\section{Systematic Uncertainties}
\label{Systematics}
We have also performed a study of the uncertainties that contribute to
the systematic errors of MAGIC. Different aspects of the imaging air
Cherenkov technique have been considered to estimate their
effect on the absolute energy scale, the gamma-ray efficiency and
the reconstructed spectra. The list of elements contributing to the
final systematic errors is long and the details will be available
in~\cite{JSitarek2011}. The overall effect is estimated to be an
uncertainty in the energy scale of 17\% at low energies and of 15\% at
medium energies. The systematic error on the slope of the measured
Crab spectrum is estimated to be 0.15, much higher than the
statistical error of the analysed data sample. The additional error on the flux
normalisation is estimated to be 19\% at low energies and 11\% at
medium energies. These errors are in agreement with the differences in
the range of 20-30\% observed between the MAGIC Crab Nebula spectrum
and the spectra measured by other experiments.



\clearpage

\end{document}